\documentclass[aps,preprintnumbers,nofootinbib,superscriptaddress,twocolumn]{revtex4}
\usepackage{epsfig,graphics}
\usepackage{amsfonts,amssymb,amsmath}            
\usepackage{pifont}                              
\usepackage{ifthen}
\usepackage{subfigure}
\usepackage{multirow}                           

\newcommand{\comment}[1]{}

\begin{document}

\title{Defining the local part of a hidden variable model: a comment}

\date[]{5th February 2010}

\author{Roger \surname{Colbeck}}
\email[]{colbeck@phys.ethz.ch}
\affiliation{Institute for Theoretical Physics, ETH Zurich, 8093
 Zurich, Switzerland.}
\affiliation{Institute of Theoretical Computer Science, ETH Zurich, 8092
 Zurich, Switzerland.}
\author{Renato \surname{Renner}}
\email[]{renner@phys.ethz.ch}
\affiliation{Institute for Theoretical Physics, ETH Zurich, 8093
 Zurich, Switzerland.}

\begin{abstract}
  In [Physical Review Letters {\bf 101}, 050403 (2008)], we showed
  that quantum theory cannot be explained by a hidden variable model
  with a non-trivial local part.  The purpose of this comment is to
  clarify our notion of \emph{local part}, which seems to have caused
  some confusion in the recent literature.  This notion is based on
  Bell's and demands that local hidden variables are \emph{physical},
  the idea being that, if discovered, they would not contradict basic
  physical principles.  We explain why the recent supposed
  ``counterexamples'' that have appeared are not counterexamples to
  our theorem|in fact they are based on a definition of local hidden
  variables which would allow signalling and is therefore not
  physical.
\end{abstract}

\maketitle

\emph{Introduction.}|In a famous paper, John Bell asked the question
as to whether there could be hidden (as yet undiscovered) parameters
that determine the seemingly random outcomes of quantum
experiments~\cite{Bell}.  Such parameters he termed \emph{local hidden
  variables}.  Local hidden variables are \emph{physical} in the sense
that, if they were discovered, they would not contradict the
no-signalling principle.\footnote{To quote Bell~\cite{Bell},
  \emph{locality} is the requirement that if a theory is supplemented
  by additional variables then ``...the result of a measurement on one
  system [is] unaffected by operations on a distant system with which
  it has interacted in the past...'', i.e.\ operations on separated
  systems cannot be used to signal.  Therefore, local hidden variables
  might alternatively be called \emph{non-signalling hidden
    variables}.}  Bell's theorem shows that local hidden variables
cannot \emph{completely} determine the experimental outcomes.

In a recent Letter~\cite{ColbeckRenner}, we asked the question as to
whether local hidden variables can betray \emph{some} information
about the outcomes.  Our main result is that they cannot.  For
example, for quantum measurements that give equally likely outcomes,
there cannot exist undiscovered observables that provide any
indication about which outcome is more likely.  This result has
sparked some controversy (see the articles by
Wechsler~\cite{crit0,crit1} and Larsson and
Cabello~\cite{crit2v1,crit2v2,crit2v3,crit2v4}\footnote{Note that the
  alternative versions of this paper each contain different arguments,
  to which we respond individually below.}), at the heart of which is
the notion of local hidden variables.  The notion we use is based on
Bell's and is physical in the sense above.  We give its mathematical
definition below.\bigskip

\begin{figure} 
\includegraphics[width=0.45\textwidth]{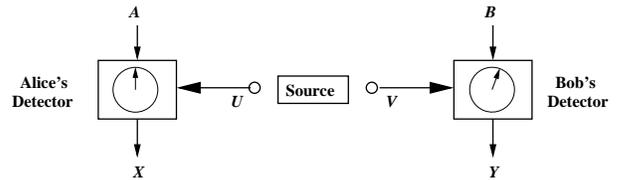}
\caption{A source emits two particles which travel to distant
  detectors.  The particles carry local hidden variables, $U$ and $V$.
  Alice can choose measurement $A$ (likewise Bob $B$) and receives
  output $X$ (respectively $Y$).}
\label{fig:hv}
\end{figure}

\emph{The notion of local hidden variables.}|We use the notation of
our original Letter~\cite{ColbeckRenner} which we summarize here (see
also Figure~\ref{fig:hv}).  Consider a source emitting two particles,
which travel to two detectors controlled by Alice and Bob.  We assume
Alice and Bob are free to choose their measurement settings, which we
denote by $A$ and $B$.\footnote{\label{ft:3}The assumption of freedom
  of choice of a measurement setting, $A$, implies in particular that,
  for any pre-existing data, $\Theta$ (which could include the
  settings and outcomes of measurements already made), the
  distribution $P_{A\Theta}$ can be chosen to be product, i.e.\ of the
  form $P_A\times P_\Theta$.}  The measurement devices generate the
outcomes $X$ and $Y$ on Alice's and Bob's sides respectively.  We
introduce \emph{local hidden variables}, denoted $U$ and $V$, with an
arbitrary joint distribution, $P_{UV}$.  These additional variables
are required to be physical in the sense that they would not enable
signalling between Alice and Bob, i.e.\ the relations
\begin{eqnarray}\label{eq:1} 
  P_{X|ABUV}=P_{X|AUV}\,\text{
    and }\, P_{Y|ABUV}=P_{Y|BUV}
\end{eqnarray}
are satisfied.\footnote{\label{ft:4}In \cite{ColbeckRenner}, we also
  introduced \emph{non-local} hidden variables as ones that are not
  physical in the above sense.  These are not needed for our
  statements, which are only about the marginal distribution
  $P_{XY|abuv}$.  (NB.\ the formula for $P_{XY|abuv}$ given in
  \cite{ColbeckRenner} is for the case where there is a non-local
  hidden variable, $W$, which is independent of $U$ and $V$.  More
  generally, $P_{XY|abuv}:=\sum_w P_{W|uv}(w)P_{XY|abuvw}$.)}  In
addition, $U$ and $V$ are said to be \emph{trivial} if
$P_{X|AUV}=P_{X|A}$ and $P_{Y|BUV}=P_{Y|B}$ (i.e.\ if they do not
convey any information about the outcomes, $X$ and $Y$).\footnote{Note
  that the treatment here is more general than that
  in~\cite{ColbeckRenner}, which is recovered in the case where $U$
  contains a copy of $V$ and vice-versa.}  By way of illustration, we
give an explicit hidden variable model in Appendix~A.

Note that hidden variables which are not local according
to~\eqref{eq:1} necessarily violate the assumption that Alice and Bob
are free to choose their measurement settings.  To see this, assume
$P_{X|ABUV}\neq P_{X|AUV}$ and that Alice makes her measurement first.
It follows that $P_{BXAUV}\neq P_B\times P_{XAUV}$, and hence that
Bob's choice of measurement, $B$, is not free (cf.\
Footnote~\ref{ft:3}).

The main result of~\cite{ColbeckRenner} is that any local hidden
variables compatible with quantum theory are necessarily trivial.  We
proceed by discussing the recent criticisms of this result. \bigskip

\begin{table}[!t]
\begin{center}
\begin{tabular}{c|cccc|}
&\multicolumn{4}{|c|}{$(X,Y)$}\\
$(A,B)$&\;$(+1,+1)$\;&\;$(+1,-1)$\;&\;$(-1,+1)$\;&\;$(-1,-1)$\;\\
\hline
$\neq(0,3)$&$\frac{1}{2}-\alpha$&$\alpha$&$\alpha$&$\frac{1}{2}-\alpha$\\
$(0,3)$&$\alpha$&$\frac{1}{2}-\alpha$&$\frac{1}{2}-\alpha$&$\alpha$\\
\hline
\end{tabular}
\caption{Probability distribution $P_{XY|AB}$ for the CHSH
  correlations ($\alpha=\frac{1}{2}\sin^2\frac{\pi}{8}$).}
\label{tab:1}
\end{center}

\begin{center}
\begin{tabular}{cc|cccc|}
&&\multicolumn{4}{|c|}{$(X,Y)$}\\
$(U,V)$&$(A,B)$&\;$(+1,+1)$\;&\;$(+1,-1)$\;&\;$(-1,+1)$\;&\;$(-1,-1)$\;\\
\hline
$(0,0)$&$\neq(0,3)$&$\frac{1}{2}$&0&0&$\frac{1}{2}$\\
$(0,0)$&$(0,3)$&0&$\frac{1}{2}$&$\frac{1}{2}$&0\\
$(+1,+1)$&all&1&0&0&0\\
$(+1,-1)$&all&0&1&0&0\\
$(-1,+1)$&all&0&0&1&0\\
$(-1,-1)$&all&0&0&0&1\\
\hline
\end{tabular}
\caption{Probability distribution $P_{XY|ABUV}$ for the hidden variable
  model described in Appendix~A.  One chooses $(U,V)=(0,0)$ with
  probability $1-4\alpha$ and $(U,V)=(\pm 1,\pm 1)$ with probability
  $\alpha$ to recreate the correlations given in Table~\ref{tab:1}.}
\label{tab:2}
\end{center}
\end{table}

\emph{Response to critics.}|In~\cite{crit1}, it is explicit that the
hidden variables considered do not satisfy the no-signalling principle
and so are not local according to our definition.  They are therefore
not physical in the sense described previously.

\comment{could not exist as part of the theory without violating current
physical laws.  

This is in contrast to our definition, where local
hidden variables (while undiscovered) are in principle discoverable
within the present theory.}

In~\cite{crit2v1}, it is argued that our framework is not general
enough to model signalling distributions.  We give an example which
shows that this is not the case in Appendix~B.  This criticism
reappeared in~\cite{crit2v4}.  Our counterargument remains the same.

In the explicit model given in~\cite{crit2v1}, if Alice were to learn
$U$, Bob could signal to her.  This is apparent from Eqn.~(12)
in~\cite{crit2v1} where it is shown that $P_{X|abu}$ is not
independent of $b$.  The hidden variable $U$ is therefore not local
according to our definition (and in particular would violate the
assumption that Bob can choose his measurement freely, as explained
above).

In~\cite{crit2v2}, a modified model is proposed.  As Eqn.~(9)
of~\cite{crit2v2} shows, the local hidden variables in this model do
not give any information about the measurement outcomes and hence are
trivial, in contrast to the authors' claim.  Their model is hence in
direct agreement with our result~\cite{ColbeckRenner}.

In~\cite{crit2v3}, a theorem is introduced which is intended to show
that even classical correlations do not fit into our model.  However,
this theorem is incorrect: the flaw in the proof is the use of the
undeclared (and unphysical) assumption that the hidden variables are
completely uncorrelated, i.e.\ that $P_{UV}=P_U\times P_V$.  Moreover,
it is easy to see that all classical correlations fit into our
model.

In~\cite{crit2v4}, the incorrect theorem of~\cite{crit2v3} is removed
and a fresh criticism is presented.  We disagree with the new
criticism and point out several problems with it below.

First, a new definition of local is used, which is not physical in the
sense described in the introduction.  In particular, the equality
$P_{AYBUV} = P_A\times P_{YBUV}$ would not hold, which means that
Alice would no longer be able to choose her measurement setting, $A$,
freely.

In addition, Larsson and Cabello claim that our definition of local
places an additional non-signalling restriction on the non-local part.
This is not the case: as we explain in Footnote~\ref{ft:4}, our result
can be formulated without mention of non-local hidden variables, which
are hence not restricted in any way.

Furthermore, the authors use their Eqn.~(6) (which states that
$P_{X|AB}=P_{X|A}$) to justify this claim.  We emphasize that this
relation does not have to hold to model correlations within our
framework (see also Appendix~B).  However, since all quantum
correlations obey this relation, Larsson and Cabello's reasoning would
lead to the conclusion that non-local hidden variables cannot exist
for any quantum correlations.  This conclusion is not correct: the
fact that $P_{X|AB}=P_{X|A}$ does not imply the impossibility of
adding a non-local hidden variable $W$ which enables signalling.

\bigskip

\emph{Appendix A: A restricted theory with a non-trivial local
  part.}|To illustrate the meaning of a non-trivial local part, we
consider a restricted theory in which there are only two devices, each
of which can make two possible measurements.  This theory can be
explained by a hidden variable model with a non-trivial local part.
Note that, because it does not contain the full set of quantum
correlations, this model is not in contradiction with the main claim
in~\cite{ColbeckRenner}.

We label the measurements $A\in\{0,2\}$ and $B\in\{1,3\}$
(cf.~\cite{ColbeckRenner}), and the outputs $X,Y\in\{+1,-1\}$. The
correlations are always such that they maximally violate the CHSH
inequality~\cite{CHSH} and are given in Table~\ref{tab:1}, where we
define $\alpha:=\frac{1}{2}\sin^2\frac{\pi}{8}$.

We can then define local hidden variables $(U,V)$ as in
Table~\ref{tab:2}.  By choosing $(U,V)=(0,0)$ with probability
$1-4\alpha$ and $(U,V)=(\pm 1,\pm 1)$ with probability $\alpha$, this
hidden variable model perfectly recreates the specified correlations.
Furthermore, the distribution $P_{XY|ABUV}$ is non-signalling and if
$U$ ($V$) was measurable, $X$ ($Y$) could be guessed correctly with
probability $\frac{1}{2}+2\alpha$.  Hence the hidden variable model
given for these correlations has a non-trivial local part. \bigskip

\emph{Appendix B: Signalling correlations}|Theorem 1 of~\cite{crit2v1}
is intended to show that our framework is not general enough to
explain signalling correlations.  In fact, the completely signalling
correlations $X=B$, $Y=A$ fit trivially into our framework.  Since it
is impossible to add variables to this model such that the resulting
distribution is non-signalling, no model with local hidden variables
can exist for any signalling correlations.

\end{document}